# An optimized pipeline for functional connectivity analysis in the rat brain

Yujian Diao[1], Ting Yin[2], Rolf Gruetter[1] and Ileana O. Jelescu[2,*]

[1]Laboratoire d'Imagerie Fonctionnelle et Métabolique, EPFL, Lausanne, Switzerland
[2]Centre d'Imagerie Biomédicale, EPFL, Lausanne, Switzerland

**\* Correspondence:**
Dr. Ileana Jelescu
ileana.jelescu@epfl.ch

**Abstract**

Resting state functional MRI (rs-fMRI) is a widespread and powerful tool for investigating functional connectivity and brain disorders. However, functional connectivity analysis can be seriously affected by random and structured noise from non-neural sources such as physiology. Thus, it is essential to first reduce thermal noise and then correctly identify and remove non-neural artefacts from rs-fMRI signals through optimized data processing methods. However, existing tools that correct for these effects have been developed for human brain and are not readily transposable to rat data. Therefore, the aim of the present study was to establish a data processing pipeline that can robustly remove random and structured noise from rat rs-fMRI data. It includes a novel denoising approach based on the Marchenko-Pastur Principle Component Analysis (MP-PCA) method, FMRIB's ICA-based Xnoiseifier (FIX) for automatic artefact classification and cleaning, and global signal regression. Our results show that: I) MP-PCA denoising substantially improves the temporal signal-to-noise ratio; II) the pre-trained FIX classifier achieves a high accuracy in artefact classification; III) both artefact cleaning and global signal regression are essential steps in minimizing the within-group variability in control animals and identifying functional connectivity changes in a rat model of sporadic Alzheimer's disease, as compared to controls.

## 1    Introduction

Resting-state fMRI (rs-fMRI) based on spontaneous low-frequency fluctuations in the blood oxygen level dependent (BOLD) signal in the resting brain is a widely used non-invasive tool for studying intrinsic functional organization in health and disease (Fornito and Bullmore, 2010; Fox and Raichle, 2007). By examining spatio-temporal correlations of the BOLD signal between distinct brain regions, known as functional connectivity, this technique is capable of revealing large-scale resting state networks (RSNs) (Biswal et al., 1995; Buckner et al., 2013; Damoiseaux et al., 2006). Nowadays, rs-fMRI has become an increasingly important translational neuroimaging tool for understanding neurological and psychiatric diseases and for developing treatments, with rapidly growing applications not only in human research but also in rodent models of disease (Bajic et al., 2017; Fox and Greicius, 2010).

However, the BOLD signal is contaminated by multiple physiological and non-physiological sources of noise, such as respiratory and cardiac cycles, thermal noise, changes in blood pressure, and head motion (Birn, 2012; Kruger and Glover, 2001; Murphy et al., 2013; Van Dijk et al., 2012). These non-neuronal sources can severely affect rs-fMRI time series and thereby confound the connectivity analysis (Cole et al., 2010; Power et al., 2014). Therefore, a robust pre-processing pipeline is required to extract the neuronal component of BOLD signal and minimize the contribution of such noise sources. Furthermore, existing tools that correct for the effect of non-neuronal sources are mostly tailored for human rs-fMRI data and are not readily transposable, or even applicable, to rodent data. Dedicated pipelines for rodent rs-fMRI processing are just starting to emerge (Bajic et al., 2017; Zerbi et al., 2015).

For example, model-based approaches such as the general linear model (GLM) can estimate and remove signal fluctuations resulting from respiratory and cardiac cycles by recording the physiology and modelling these external confounds as regressors (Birn et al., 2006; Kasper et al., 2017). While physiological recordings in rodents are possible, they typically involve dedicated hardware and invasive procedures, making them experimentally difficult. However, although cardiac and respiratory frequencies in rodents are much higher than those of the resting-state BOLD fluctuations, depending on the temporal resolution of the acquisition, they can alias into the band of interest (typically $0.01 – 0.3$ Hz) and corrupt the analysis. Two complementary approaches are, therefore, suitable to mitigate the impact of physiological noise in resting-state rodent fMRI.

One approach is the removal of global signal defined as the mean time series averaged over all voxels within the brain by including the global signal as a nuisance regressor in GLM analyses, which is referred to as global signal regression (GSR) (Liu et al., 2017). However, the use of GSR has been one of the most controversial topics in human rs-fMRI connectivity studies (Liu et al., 2017; Murphy and Fox, 2017). On one hand, GSR is known to introduce spurious negative correlations (Murphy et al., 2009) and cause spatial bias on connectivity measures (Saad et al., 2012). On the other hand, prior studies have shown that GSR can enhance the detection of significant functional connectivity and improve spatial specificity of positive correlations (Fox et al., 2009). Most importantly for rodent studies, GSR can also mitigate confounds related to motion and physiological processes (Aquino et al., 2019; Power et al., 2015).

Another commonly used data-driven approach which identifies various physiological noise components directly from the fMRI data itself is single-level independent component analysis (ICA) (Bajic et al., 2017; Caballero-Gaudes and Reynolds, 2017; Griffanti et al., 2014; McKeown et al., 2003). The ICA method is also confronted by several issues including model order selection(i.e. the number of components) (Kuang et al., 2018) and the identification of artefactual components, which is a manually tedious step (Wang and Li, 2015), especially for a high order model. Notably, a machine-learning approach for automatic artefact component classification based on FMRIB's ICA-based Xnoiseifier (FIX) (Salimi-Khorshidi et al., 2014) has been proposed to replace manual classification. The FIX auto-classifer applied in human and mice rs-fMRI studies has yielded promising results with a high accuracy in artefact identification (Griffanti et al., 2015, 2014; Zerbi et al., 2015). However, the success of FIX classification relies on a proper pre-training on study-specific datasets.

Therefore, the aim of the present study was to propose and evaluate a data processing pipeline for rat rs-fMRI that minimizes intra-group variability and maximizes between-group differences in whole-brain functional connectivity. In this pipeline, we reduced structural noise by combining single-session ICA cleaning and GSR. For ICA cleaning, we built and used a dedicated FIX classifier for rats. Furthermore, we also enhanced the sensitivity to BOLD fluctuations by first increasing dramatically the temporal SNR of the data. For the purpose of stochastic (thermal) noise removal, we employed a novel method based on Marchenko-Pastur Principle Component Analysis (MP-PCA). MP-PCA denoising was recently introduced for diffusion MRI and is a model-free method that exploits redundancy in MRI series (Veraart et al., 2016), which has shown great potential for improving the SNR in other MRI techniques as well (Ades-Aron et al., 2019, 2018; Does et al., 2019).

## 2   Materials and Methods

### 2.1   Animal preparation and anesthesia

All experiments were approved by the local Service for Veterinary Affairs. Male Wistar rats (236±11 g) underwent a bilateral icv-injection of either streptozotocin (3 mg/kg, STZ group) or buffer (control group). When delivered exclusively to the brain, streptozotocin induces impaired brain glucose metabolism and is used as a model of sporadic Alzheimer's disease (Grieb, 2016; Knezovic et al., 2015; Lester-Coll et al., 2006).

The fMRI data were acquired on 17 rats in two groups, disease group (STZ, 9 rats) and control group (CTL, 8 rats), at 2, 6, 13 and 21 weeks after the surgery, illustrated in **Figure 1**. Rats were anesthetized using 2% isoflurane in a mixture of $O_2$ and Air ($O_2$/Air: 30/70) during initial setup and promptly switched to medetomidine sedation delivered through a subcutaneous catheter in the back (bolus: 0.1mg/kg, perfusion: 0.1mg/kg/h) as previously described (Reynaud et al., 2019). Medetomidine preserves neural activity and vascular response better than isoflurane (Grandjean et al., 2014; Pawela et al., 2009; Weber et al., 2006). The rat head was fixed using a homemade holder with a bite bar and ear bars to minimize the head motion, and body temperature and breathing rate were continuously monitored. At the end of the scanning sessions, rats were woken up with an intra-muscular injection of antagonist atipamezole (0.5 mg/kg) and returned to their cages.

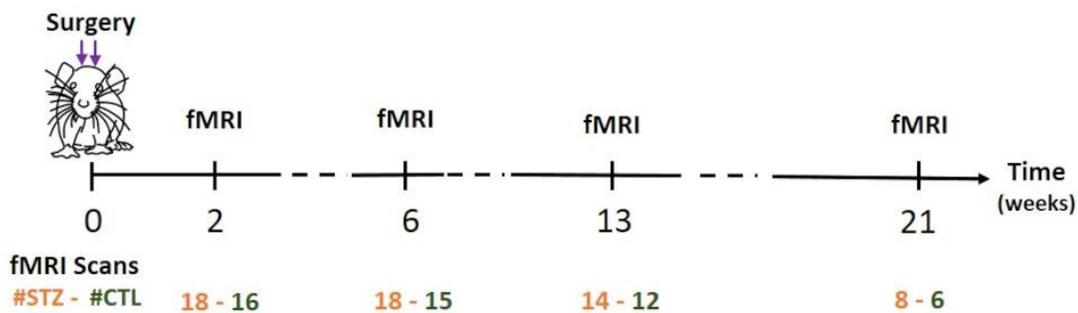

**Figure 1.** Time line of experiments. Two fMRI runs were acquired per rat for each experiment.

## 2.2 MRI acquisition

MRI experiments were conducted on a 14.1 T small animal Varian system (Varian Inc.) equipped with 400 mT/m gradients, using an in-house built quadrature surface transceiver.

An anatomical reference scan was acquired using a fast spin-echo multi-slice sequence with following parameters: TE/TR = 10.17/3000 ms, ETL = 4, $TE_{eff}$ = 10.17 ms, field of view (FOV) = 19.2 × 19.2 $mm^2$, matrix = 128 × 128, in-plane resolution = 150 × 150 $\mu m^2$, number of slices = 30, thickness = 0.5 mm. Before running the fMRI sequence, anesthesia was switched from isoflurane to medetomidine. The fMRI acquisitions were started after a fixed duration (~ 1 hour) since the switch from isoflurane to medetomidine to minimize between-animal anesthesia-related confounds. Rs-fMRI data were acquired using a two-shot gradient-echo EPI sequence as follows: TE = 10 ms, TR = 800 ms, $TR_{vol}$ = 1.6 s, FOV = 23 × 23 $mm^2$, matrix = 64 × 64, in-plane resolution = 360 × 360 $\mu m^2$, 8 slices, thickness = 1.12 mm, 370 repetitions (~ 10 minutes). Two fMRI runs were acquired for each rat. A short scan with 10 repetitions and reversed phase-encoding direction was also acquired to correct for EPI-related geometric distortions.

## 2.3 FMRI data pre-processing

Images were first skull-stripped automatically using BET (Brain Extraction Tool; FSL, https://fsl.fmrib.ox.ac.uk/fsl) (Smith, 2002) and fMRI timeseries were denoised using MP-PCA with a 5×5×5 voxel sliding kernel (Veraart et al., 2016). The quality of MP-PCA denoising was assessed by inspecting the normality of the residuals (original - denoised) and the temporal signal-to-noise ratio (tSNR) changes before and after denoising. Specifically, the normality of the residuals was tested by the linearity of the relationship between the natural logarithm of the residual distribution probability and the squares of multiple residual standard deviation. Then, the datasets went through EPI distortion correction using FSL's topup (Smith et al., 2004), slice-timing correction (Calhoun et al., 2000; Henson et al., 1999; Sladky et al., 2011), and spatial smoothing (Gaussian kernel: 0.36×0.36×1 mm$^3$). Corrected fMRI images were registered to the Waxholm Space Atlas of the rat brain (https://www.nitrc.org/projects/whs-sd-atlas) using linear and non-linear registration in ANTs (Avants et al., 2008) and 28 atlas-defined ROIs (14 per hemisphere) were automatically segmented.

Finally, single-session ICA was performed on fMRI timecourses using FSL's MELODIC (Beckmann and Smith, 2004) with high-pass temporal filtering ($f > 0.01$ Hz) and 40 independent components (IC's).

## 2.4 FIX training

Datasets were randomly split into two groups: a training dataset for FIX (n = 49) and a test dataset (n = 58). The ICA components in the training dataset were manually classified to signal or artefact, which was mainly based on thresholded spatial maps because ICA is theoretically more robust in the spatial than in the temporal domain (Salimi-Khorshidi et al., 2014; Smith et al., 2012). Here we chose an "aggressive" artefact removal (Griffanti et al., 2014) in the training dataset in order to give the trained classifier a margin to be conservative or aggressive via adjusting the threshold fed to it (small thresholds make it conservative).

The performance of the trained classifier in detecting artefactual components was evaluated on the test dataset by comparing the automatic classification of artefactual components with the manual classification. The classification accuracy was characterized in terms of "recall" and "precision" (Powers, 2011), which are defined as the percentage of the correctly predicted artefact components in all actual artefact components and the percentage of the correctly predicted artefact components in all predicted artefact components, respectively.

## 2.5 Network analysis and global signal regression

After ICA decomposition and classification, the artefact components were regressed out of the 4D pre-processed datasets to obtain "cleaned" rs-fMRI datasets. The cleaned data were used to compute ROI-to-ROI functional connectivity by calculating correlation coefficients between the ROI-averaged time-series of the 28 atlas-defined ROIs, resulting in a 28 × 28 FC matrix for each rat.

For FIX, we relied on spatial maps rather than timecourses to evaluate artefactual components. In several cases, components displayed sensible spatial distribution to represent a RSN but the power spectrum showed a peak at a frequency which could be attributed to breathing (**Figure 2**) (the breathing rate was recorded for each run and its aliased frequency within our 0.01 – 0.31 Hz band calculated). These non-neuronal sources which could not be cleaned using FIX also contribute to the global signal. To mitigate their effect, we used partial correlation of ROI-to-ROI timecourses to build FC matrices, with the global signal as the controlling variable. For every pair of ROIs, the partial correlation was implemented by measuring the correlation between their time-series residuals, after each having been adjusted by the global signal regression (Smith et al., 2011).

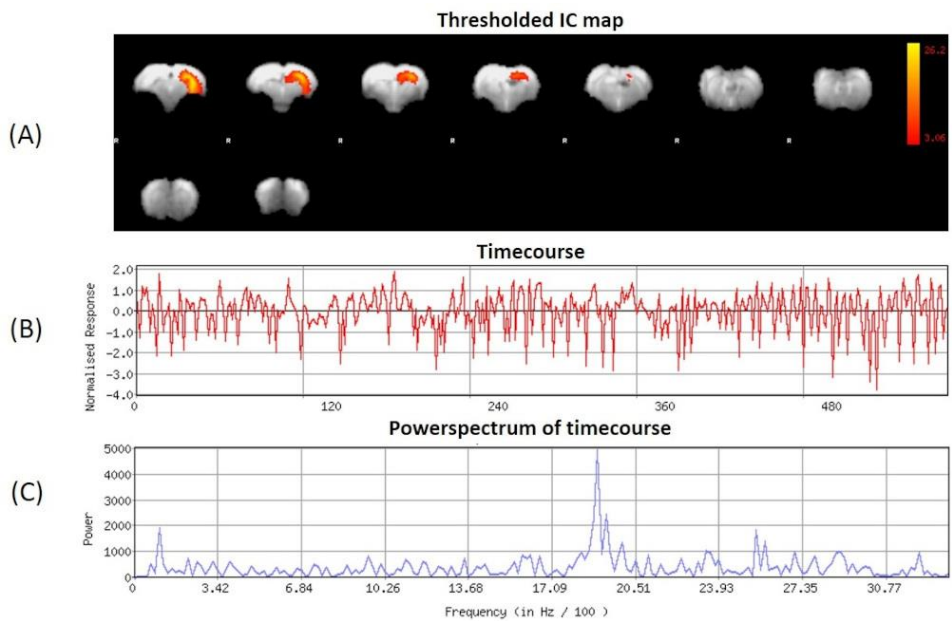

**Figure 2.** Example of independent component which is anatomically consistent (left hippocampus) yet corrupted by breathing: thresholded spatial map (**A**), timecourse (**B**) and the power spectrum of timecourse (**C**). The power spectrum shows an aliased frequency with a peak at around 0.18 Hz due to breathing.

Finally, statistical comparisons of functional connectivity between STZ and CTL groups at each timepoint were performed using NBS (Zalesky et al., 2010) to identify network connections that showed significant between-group difference. Specifically, NBS uses one-tailed two-sample *t*-test to detect differences in group-averaged FC between the two groups. Thereby, two contrasts (*STZ>CTL* and *STZ<CTL*) were tested separately. A *t*-statistic threshold was chosen on the basis of medium-to-large sizes of the subnetwork comprised of connections with their *t*-statistic above the threshold (Tsurugizawa et al., 2019) as well as the underlying *p*-values. Here, we chose 2.2 as the *t*-statistic threshold. Significance ($p \leq 0.05$) was tested after family-wise error rate (FWER) correction using non-parametric permutation ($N = 5000$).

The full data processing pipeline is illustrated in **Figure 3**.

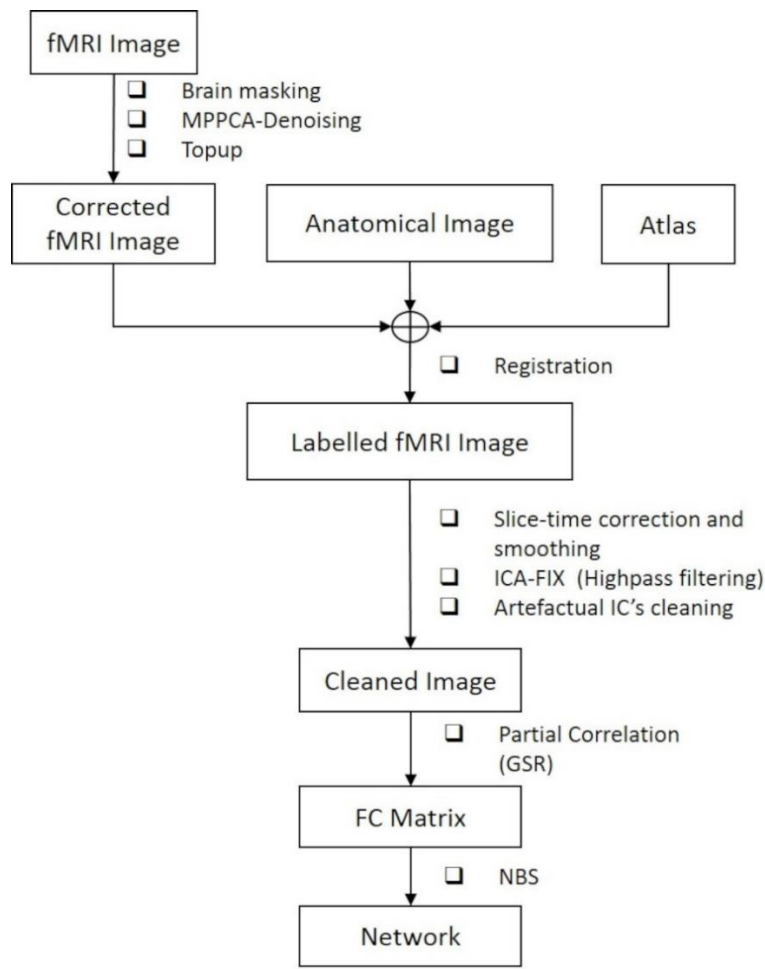

**Figure 3.** The proposed pipeline for rs-fMRI data processing.

## 2.6 Data processing pipeline evaluation

To evaluate the rs-fMRI data processing pipeline including MP-PCA denoising (DN), slice-timing correction (SC), spatial smoothing (SM), high-pass filtering (HP), ICA-FIX cleaning (CL) and global signal regression (GSR) in terms of consistency of within-group functional connectivity in healthy CTL group and in terms of between-group difference, we compared results of the optimized pipeline (DN + SC + SM + HP + CL + GSR, D) with three other processing approaches excluding ICA-FIX cleaning and/or GSR, namely, DN + SC + SM + HP (A), DN + SC + SM + HP + CL (B), and DN + SC + SM + HP + GSR (C), shown in **Table 1**.

**Table 1.** The four data processing pipelines and methods they include ("×" – including, "○" – excluding)

| Pipelines | MPPCA-denoising (DN) | Slice-timing correction (SC) | Spatial smoothing (SM) | High-pass filtering (HP) | ICA-FIX cleaning (CL) | Global signal regression (GSR) |
|---|---|---|---|---|---|---|
| A | × | × | × | × | ○ | ○ |
| B | × | × | × | × | × | ○ |
| C | × | × | × | × | ○ | × |
| D | × | × | × | × | × | × |

Based on the hypothesis that an optimal processing procedure should minimize the variability within the homogeneous group of healthy controls (Zerbi et al., 2015), the within-group variability was assessed by calculating the standard deviation of the Fisher z-transformed correlation coefficients of the FC matrices in the CTL group of healthy rats at each timepoint. In addition, the sensitivity to between-group differences was evaluated by comparing the significant difference in FC between STZ and CTL group at each timepoint.

## 2.7 Group ICA analysis

At each timepoint, group-level ICA was performed using FSL Melodic (Smith et al., 2004) on the pooled CTL and STZ datasets which were pre-processed using pipeline B (no GSR). Prior to group-level ICA in Melodic, all the rs-fMRI datasets from a given timepoint were registered to a common template using ANTs (Avants et al., 2008), and the template registration in Melodic was by-passed. Thirty group-level spatial IC's were extracted and dual regression (Filippini et al., 2009) was used to estimate subject-specific timecourses and associated spatial maps. Similar to the ROI-based FC analysis, ICA-based functional connectivity matrices were built by calculating correlation coefficients between all pairs of timecourses (excluding artefact components) using FSLNets (Griffanti et al., 2014). Both full correlation (original Pearson's correlation) and partial correlation with GSR (mean timecourse as regressor) were employed by FSLNets to evaluate the similarity between timecourses.

In the end, differences in functional connectivity between STZ and CTL groups were tested using the NBS toolbox using the same parameters in Section 2.5.

## 3   Results

To experimentally evaluate the processing pipeline, a total of 109 rs-fMRI datasets were acquired from 17 rats at 4 timepoints ranging from 2 weeks to 21 weeks. Datasets with bad image quality were discarded (**Figure 1**). **Figure 4** shows the MR images in one representative dataset including rs-fMRI images, matching anatomical reference and the atlas-based anatomical labels registered to the fMRI images.

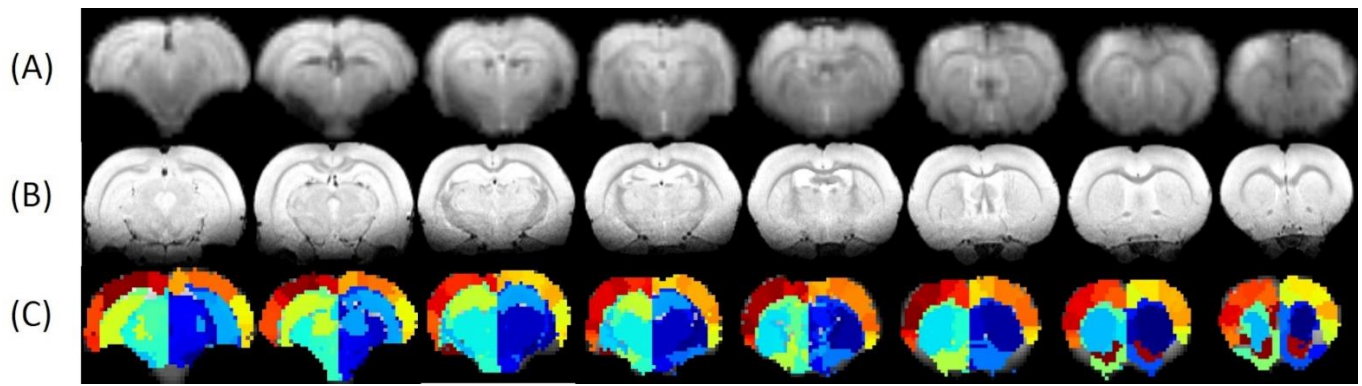

**Figure 4.** Example of rs-fMRI images of 8 coronal slices (**A**), matching anatomical MR images (**B**) and atlas-based anatomical labels registered to the fMRI images (**C**).

### 3.1   MP-PCA denoising

The average tSNR after MP-PCA denoising improved significantly for all the 107 datasets. **Figure 5** shows an example of the average tSNR increase from 75 to 146 after MP-PCA denoising. **Figure 6** shows an example of residuals map, histogram, and the normality test. The linearity of $\log(P) = f(r^2)$ confirms the residuals are normally distributed and only Gaussian noise has been removed from the signal.

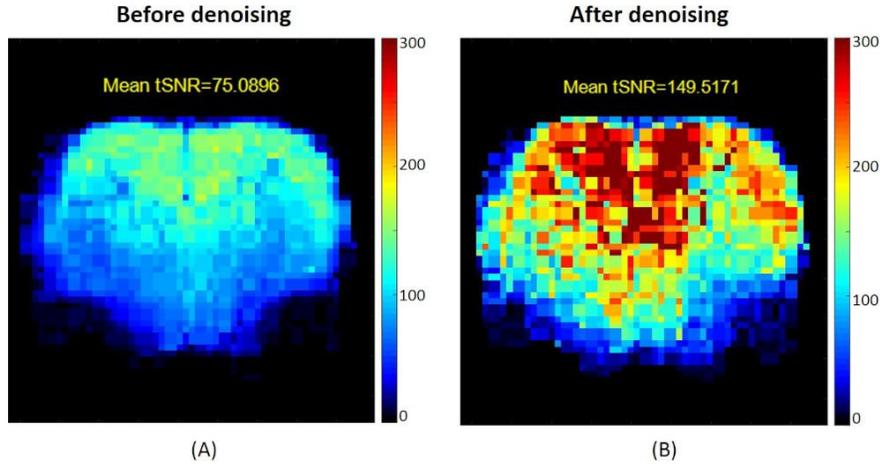

**Figure 5.** Example of temporal SNR maps, before (**A**) and after MPPCA-denoising (**B**). The mean tSNR over the middle brain slice was improved dramatically from 75 to 146. The SNR profile is typical of a surface coil placed on top of the head, with higher sensitivity in the cortex.

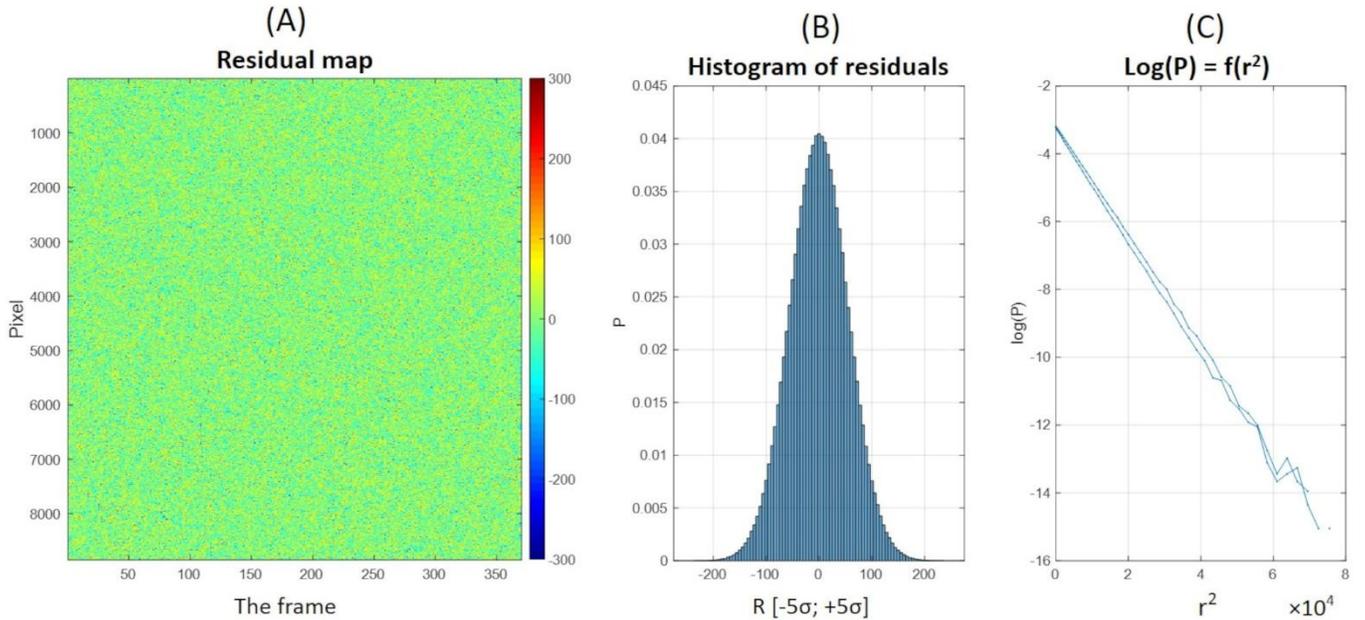

**Figure 6.** Normality estimation of the residuals after MPPCA-denoising. (**A**) Residuals map of all voxels within the brain mask (rows) and time frame (columns), (**B**) Histogram of residuals, (**C**) Normality test.

### 3.2 FIX classification

Here, we preferred a lower order model to avoid overfitting (Kuang et al., 2018; Li et al., 2007) and we chose the number of independent components to be 40, which typically explained 90% of the variance. Notably, reaching 95% of explained variance would have required about 90 components, which would potentially cause over-splitting networks and making the classification more complicated.

The single-subject ICA was performed on each dataset with 40 components. In the training dataset (n = 49), 19.8 ± 4.7 components (50 %) were classified as artefacts by hand. In the test dataset (n = 58), between 40% and 64% of components were classified as artefacts automatically by FIX depending on the threshold. More components could be recognized as artefacts by increasing the FIX threshold at the expense of lowering the classification precision due to more misclassification. Here 45 might be an "optimal" threshold with overall 88% in recall and 90% in precision achieved (**Table 2**).

**Table 2.** FIX artefact classification accuracies at different thresholds. Recall = correctly classified artefacts/all real artefacts, precision = correctly classified artefacts/all classified artefacts. As the threshold increases, both the percentage of artefacts detected by FIX and the recall increase but the precision decreases.

| FIX THRESHOLD | 20 | 30 | 40 | 45 | 50 | 60 | 70 |
|---:|---|---|---|---|---|---|---|
| Signal components (%) | 44.9 | 44.9 | 44.9 | 44.9 | 44.9 | 40.4 | 36.2 |
| Artefact components (%) | 39.6 | 43.9 | 49.4 | 52.7 | 55.1 | 59.6 | 63.8 |
| Unknown components (%) | 15.5 | 11.2 | 5.7 | 2.5 | 0 | 0 | 0 |
| FIX artefact recall (%) | 69.8 | 76.5 | 83.3 | **87.8** | 89.7 | 95.1 | 98.5 |
| FIX artefact precision (%) | 100 | 95.4 | 91.6 | **90.2** | 86.5 | 84.4 | 82.2 |

### 3.3 Comparison of pipeline performance

The four data processing pipelines were compared based on resulting functional connectivity matrices of 28 atlas-based ROIs.

Our proposed pipeline D (DN+SC+SM+HP+CL+GSR) obtained the minimal within-group variability in the homogeneous group of healthy controls for all timepoints while other procedures excluding artifact cleaning (CL) and/or GSR resulted in higher variability (**Figure 7**). Furthermore, pipeline D also yielded the most consistent and widespread between-group differences at 2, 6 and 13 weeks (**Figure 8**). No significant difference was found between CTL and STZ groups for pipeline A (without CL and GSR) and significant differences were only found at one timepoint (6 weeks) for pipeline B (CL without GSR).

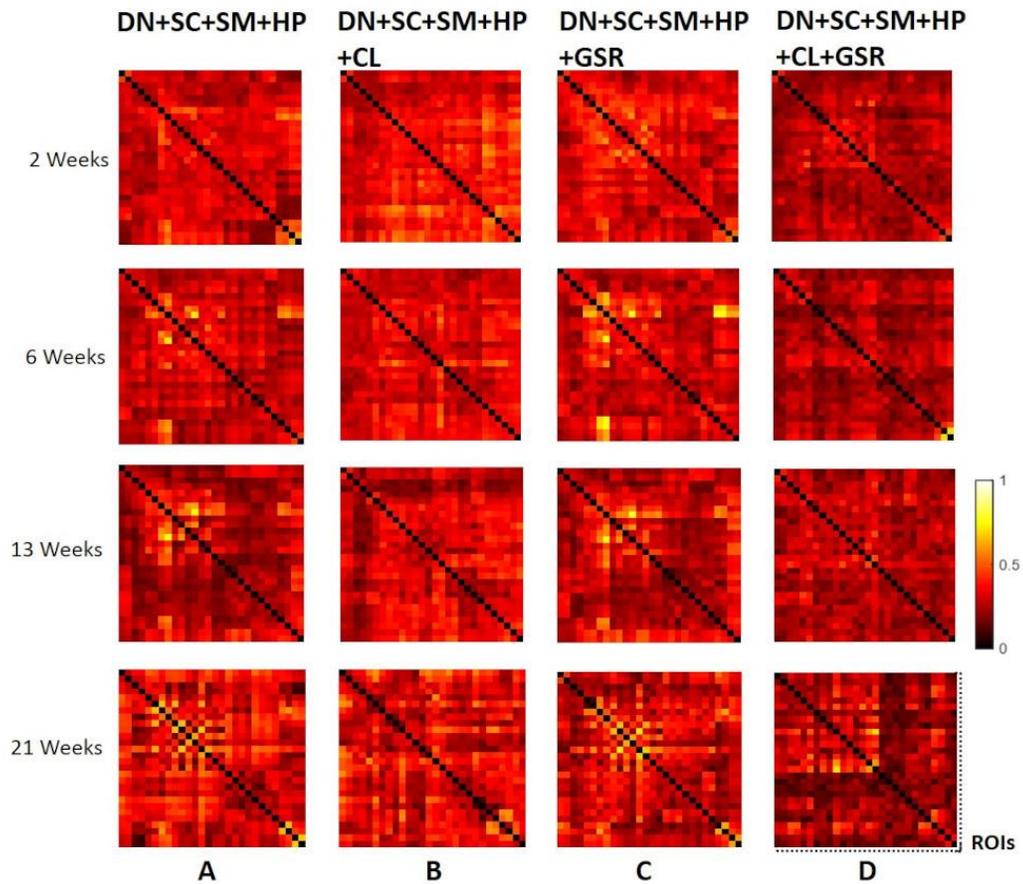

**Figure 7.** Standard deviation of the Fisher *z*-transformed correlation coefficient for functional connectivity in the CTL group at four different timepoints for the four pipelines (A-D). Our proposed pipeline D obtained the minimal within-group variability in the homogeneous group of healthy controls for all timepoints while other procedures excluding CL and/or GSR had higher variability

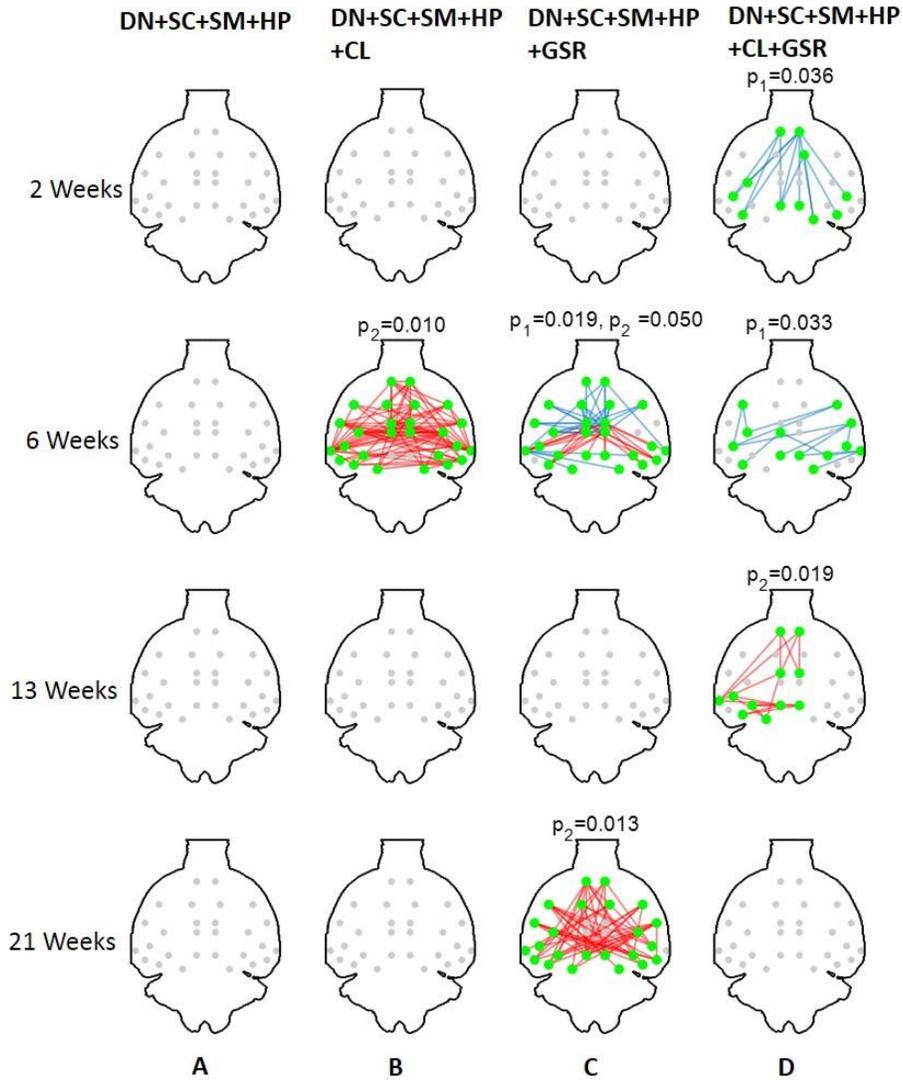

**Figure 8.** The significant difference in FC between CTL and STZ groups at each timepoint (2, 6, 13 and 21 weeks) for each data processing approach under NBS threshold 2.2. A: DN+SC+SM+HP, B: DN+SC+SM+HP+CL, C: DN+SC+SM+HP+GSR, D: DN+SC+SM+HP+CL+GSR. Pipeline D most consistently yielded between-group differences across timepoints. No significant difference was found between CTL and STZ group for pipeline A without CL and GSR while significant difference was only found at 6 weeks for pipeline B without GSR. Blue edges indicate group differences in contrast1 (*STZ > CTL*) and red edges indicate group differences in contrast2 (*STZ < CTL*). $p_1$ and $p_2$ are FWER corrected *p*-values for contrast1 and contrast2, respectively.

### 3.4 ROI-based Functional connectivity analysis

To analyze group differences in connectivity obtained following the data processing pipeline D, a more detailed visualization is provided in **Figure 9**. Brain regions with altered connectivity in STZ animals were consistent with regions affected by Alzheimer's disease and with previous findings on this animal model (Grieb, 2016; Kraska et al., 2012; Lester-Coll et al., 2006; Mayer et al., 1990; Shoham et al., 2003). At 2 weeks, increased positive connectivity of the anterior cingulate (ACC) to retrosplenial cortex (RSC), and decreased anti-correlations of default mode network (DMN) including ACC, RSC, posterior parietal cortex (PPC) and hippocampus (Hip/Sub) to lateral cortical network (LCN) involving somatosensory (S1) as well as motor (M) were found. The 6-week timepoint showed widespread reduced anti-correlations between DMN including RSC, PPC & Hip, and somatosensory of LCN, and striatum (Str). At 13 weeks, weaker positive correlations were found not only within the DMN involving ACC, PPC, RSC, Hip, medial temporal lobe (MTL) and visual cortex (V) but also hypothalamus (HTh) to DMN.

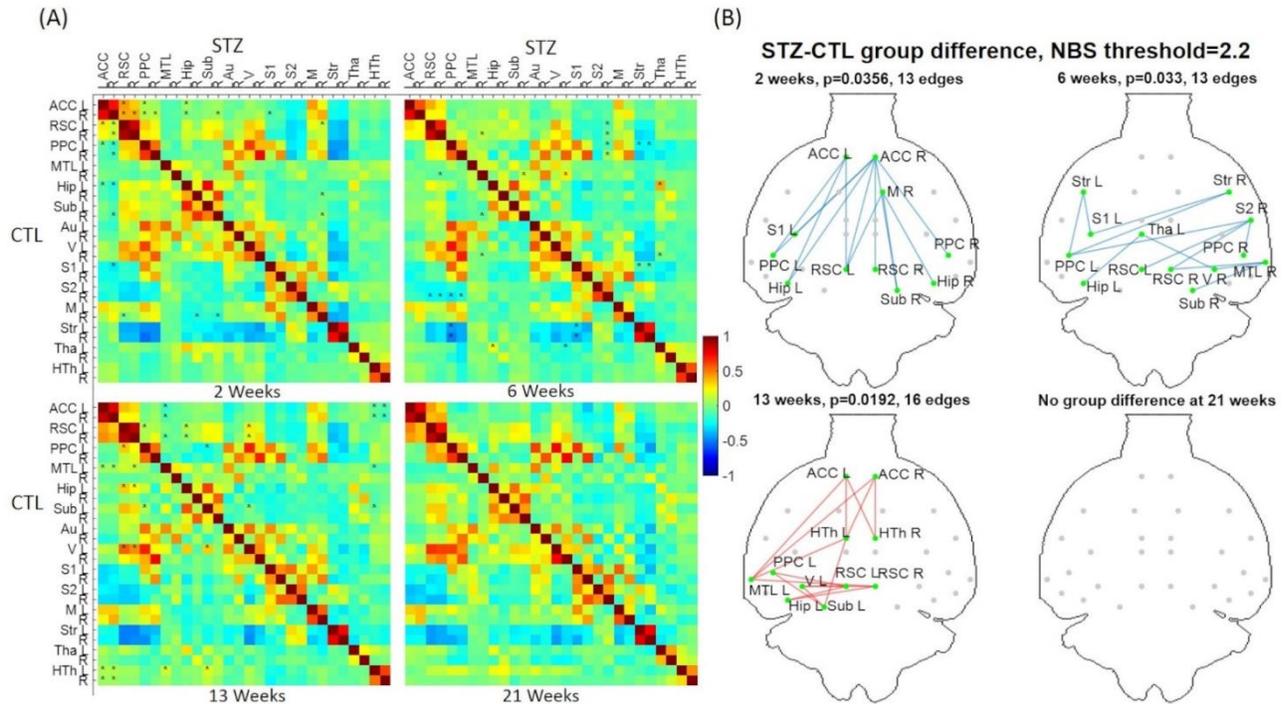

**Figure 9.** (**A**) Hybrid average FC matrices at each timepoint (top-right half: STZ, bottom-left half: CTL) based on the data processed by the optimized pipeline. *: $p < 0.05$ (FWER corrected) at threshold of 2.2. (**B**) Graph networks at 4 timepoints. Blue/red edges and green nodes indicate connections with significant difference. The *p*-value for each network was given after FWER correction. *ACC: anterior cingulate cortex; RSC: retrosplenial cortex; PPC: posterior parietal cortex; MTL: medial temporal lobe; Hip: hippocampus; Sub: subiculum; Au: auditory; V: visual; S1/S2: primary/secondary somatosensory; M: motor; Str: striatum; Tha: thalamus; HTh: hypothalamus. L/R: left/right.*

### 3.5 Group ICA

Group ICA with 30 components was carried out on the same datasets. Artefactual components were identified and removed, which resulted in 28, 25, 25 and 26 signal components left for the 4 timepoints respectively. **Figure 10** displays significant between-group differences in partial correlations with GSR between ICA components. Remarkably, group ICA analysis with GSR showed intergroup differences at 3 timepoints from 2 up to 13 weeks, in agreement with differences found in ROI-based FC analysis using Pipeline D. At 2 weeks, most changes were found in connections between RSC, PPC, Hip, thalamus (Tha) and S1/2. At 6 weeks, there were alterations found in connectivity involving

PPC, Hip, RSC, S1/2, Str, as well as Tha. At 13 weeks, primary changes were detected in connections between Hip, S1, Str, Tha and HTh. **Figure 11** shows group differences resulting from full correlation between ICA components. Notably, this analysis revealed significant differences at all timepoints, in poor agreement with the ROI-based FC analysis using Pipeline B, which only identified group differences at one timepoint (6 weeks). This inconsistency may indicate the importance of GSR in obtaining consistent intergroup differences.

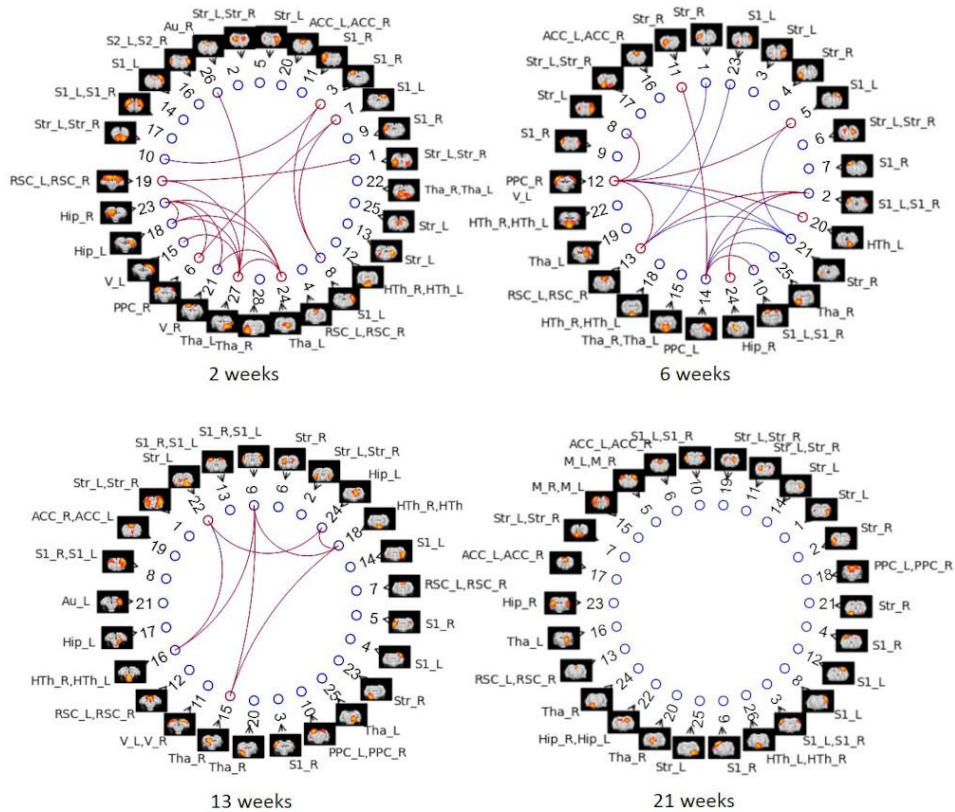

**Figure 10**. Between-group differences in ICA-based functional connectivity with GSR (partial correlation) at NBS threshold of 2.2 for each timepoint. Colored edges display the existence of STZ-CTL difference in connections between IC's. Each IC is denoted by a spatial map and its IC number. The nodes of IC's are listed in an order based on its position in brain (anterior to posterior). ROI labels are attached to every ICA component. Artefactual components were removed and the IC number was reordered accordingly. Blue edges indicate group differences in contrast1 (*STZ > CTL*) and red edges indicate group differences in contrast2 (*STZ < CTL*).

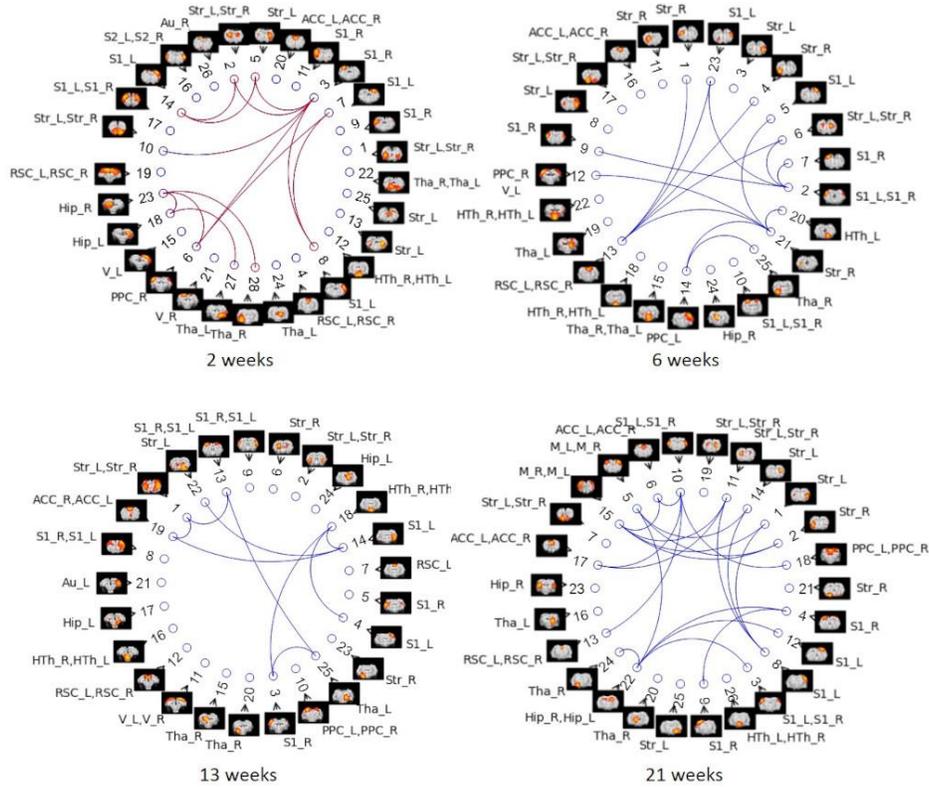

**Figure 11.** Between-group differences in ICA-based functional connectivity without GSR (full correlation) at NBS threshold of 2.2 for each timepoint. Colored edges display the existence of STZ-CTL difference in connections between IC's. Each IC is denoted by a spatial map and its IC number. The nodes of IC's are listed in an order based on its position in brain (anterior to posterior). ROI labels are attached to every ICA component. Artefactual components were removed and the IC number was reordered accordingly. Blue edges indicate group differences in contrast1 (*STZ > CTL*) and red edges indicate group differences in contrast2 (*STZ < CTL*).

## 4 Discussion

In this work, we proposed a novel resting-state fMRI processing pipeline adapted for rat data. We included a novel thermal noise reduction method based on MP-PCA applied to rs-fMRI data which substantially improved the tSNR. We also built a dedicated FIX ICA classifier for rat brain which showed a high accuracy in distinguishing artefactual ICA components from the rs-fMRI signal after training. We evaluated the performance of a pipeline that included denoising, artefact cleaning and GSR by comparing it to three other possible approaches that excluded artefact cleaning and/or GSR. We showed that these two steps were essential in minimizing the within-group variability in the healthy control group and in identifying between-group differences in functional connectivity between a control group and a diseased group using the STZ animal model. Between-group differences evaluated either via ROI-to-ROI functional connectivity or group-level ICA were more consistent between the two approaches when both cleaning and some form of GSR were included.

The MP-PCA denoising method achieved a high performance in random noise reduction. The residuals followed a Gaussian distribution, showing that MP-PCA is removing thermal noise while minimally affecting structured signal. This method resulted in substantial improvement in tSNR, which helped lay down foundations for the subsequent processing. It is often suggested that ICA also has denoising properties (Griffanti et al., 2014; McKeown et al., 2003), which needs be clarified. On one hand, if just a few artefactual components are removed from the signal (as in FIX cleaning), the effect of ICA is primarily to remove structured noise, and not thermal (random) noise. On the other hand, if the ICA decomposition is used to keep and examine just a few independent components that appear anatomically consistent with RSNs, then indeed most of thermal noise is also removed in that process (Beckmann and Smith, 2004). But group-level ICA suffers from its own limitations (Cole et al., 2010) and is not necessarily the appropriate analysis tool for all studies, and seed-based analyses are expected to benefit greatly from prior denoising using MP-PCA. Another study using MP-PCA denoising prior to task fMRI analysis reported an increase of 60% in SNR and improved statistics and extent of the activation(Ades-Aron et al., 2018).

Head motion during fMRI acquisitions is one of the major confounding factors that leads to artificial correlation compromising the interpretation of rs-fMRI data (Maknojia et al., 2019; Van Dijk et al., 2012). However, compared to human studies where head motion is common, rodent studies are less impacted by this confound due to the restraint achieved by a fixation setup with ear bars and a bite bar and the use of anesthesia (Pan et al., 2015). In our datasets, no apparent head motion was observed by visual inspection of time courses except for two datasets, which were discarded. Motion correction was therefore skipped in the proposed data processing pipeline because it has been shown to introduce spurious correlations (Chuang et al., 2019; Grootoonk et al., 2000; Sirmpilatze et al., 2019).

The FIX-based artefact auto-classification has already been applied in human and mouse fMRI datasets (Salimi-Khorshidi et al., 2014; Zerbi et al., 2015). In this work, this automatic artefact removal approach was for the first time implemented for rat data. After being trained in a manually classified dataset, the classifier showed a high accuracy in identifying artefact components from resting state fMRI signal in an untouched test dataset. Nonetheless, **Table 1** shows that there is a trade-off between the recall and precision in the fully automatic classification of artefact components, which means it is not possible to achieve both very high recall and precision with one FIX threshold. However, in practice, this problem could be addressed by half automated classification in which two auto-classifications are first performed with respectively low and high thresholds (20 and 70 for instance) and then the different components between their classified artefacts are manually examined. In this way, by examining a small portion of ICA components (~24%), we were able to achieve a very high classification accuracy in a very short time. Note that the training set was cleaned aggressively in order to give flexibility in aggressiveness/conservatism for test datasets by adjusting the threshold. This classifier is available upon request and will be uploaded to a public repository.

Interestingly, ICA-based cleaning lowered overall intra-group variability at three timepoints (6 weeks, 13 weeks and 21 weeks – Figure 7) while its effect was less systematic for datasets at 2 weeks. Pipeline B yielded significant between-group differences at 6 weeks only, as compared to the minimal pipeline A (**Figure 8**, columns A vs B). This suggests ICA-based cleaning is not sufficient for a comprehensive pre-processing pipeline of rat fMRI data.

Although controversial, global signal regression is still commonly used in the analysis of rs-fMRI data (Falahpour et al., 2018) due to its capability of reducing the effects of respiration and motion on functional connectivity estimates (Birn, 2012; Power et al., 2014; Yan et al., 2013) and enhancing the spatial specificity of positive correlations (Fox et al., 2009). Here, we found that pipeline C which included GSR had little effect in reducing intra-group variability compared to the minimal protocol A. However, combined with ICA-based cleaning, pipeline D reduced most within-group variability in the healthy CTL group and revealed most differences in connectivity between the CTL and STZ groups.

Remarkably, the combination of both cleaning and some form of GSR (be it average timecourse over the whole brain or average timecourse over all IC's) also yielded the most consistent group differences between ROI-to-ROI connectivity and ICA-based connectivity, highlighting brain regions typically affected in Alzheimer's disease and its associated animal model. Conversely, excluding GSR, ICA-based functional connectivity revealed changes in STZ rats at all timepoints while ROI-based functional connectivity only showed changes at the 6-week timepoint. This suggests that global signal regression plays an important role in functional connectivity analyses of rodent data, in terms of robustness to the choice of analysis tools and sensitivity to group differences.

## 5    Conclusion

We conclude that the processing pipeline for rat rs-fMRI data proposed herein, which includes MP-PCA denoising, a FIX auto-classification and cleaning of structured artefacts uncovered by ICA, and global signal regression, allowed to greatly reduce the within-group variability and improve the detection of between-group differences. This data processing pipeline therefore has strong potential to improve the sensitivity and reproducibility of rs-fMRI studies on rat models of disease and injury.

**Acknowledgments**

The authors thank Analina Da Silva, Mario Lepore and Stefan Mitrea for technical assistance with animal experiments. The work was supported by the Center for Biomedical Imaging of the EPFL, Unil, CHUV, UniGe and HUG.